\documentclass[letterpaper]{article} 
\usepackage[draft]{aaai2026}  
\usepackage{times}  
\usepackage{helvet}  
\usepackage{courier}  
\usepackage[hyphens]{url}  
\usepackage{graphicx} 
\usepackage{natbib}  
\usepackage{caption} 
\usepackage{algorithm}
\usepackage{algorithmic}

\usepackage{newfloat}
\usepackage{listings}
\DeclareCaptionStyle{ruled}{labelfont=normalfont,labelsep=colon,strut=off} 
\floatstyle{ruled}
\newfloat{listing}{tb}{lst}{}
\floatname{listing}{Listing}
\title{Assessing Social Alignment: Do Personality-Prompted Large Language Models Behave Like Humans?}

\author{
  \textbf{Ivan Zakazov\thanks{These authors contributed equally.}},
  \textbf{Mikolaj Boronski\footnotemark[1]},
  \textbf{Lorenzo Drudi\footnotemark[1]},
  \textbf{Robert West}\\
  \small{Correspondence: {\tt \{ivan.zakazov, mikolaj.boronski, lorenzo.drudi, robert.west\}@epfl.ch}}
}

\affiliations{
    EPFL
}

\usepackage{bibentry}

\usepackage{amsmath}
\usepackage{tabularx}
\usepackage{xurl}            

\usepackage{booktabs}        
\usepackage{tikz}            
\usetikzlibrary{shadings}    

\definecolor{gold}{RGB}{255,215,0}

\newcommand{\qualifierBox}[2]{%
  \tikz[baseline=(box.base)]{
    \node[draw=black, thick, inner sep=2pt, rounded corners, fill=#2] (box) {#1};%
  }%
}

\newcommand{\qualExtremely}{\qualifierBox{extremely}{gold!100}} 
\newcommand{\qualVery}{\qualifierBox{very}{gold!40}}           
\newcommand{\qualABit}{\qualifierBox{a bit}{gold!20}}           
\newcommand{\qualNeither}{\qualifierBox{neither}{gold!0}}      
\newcommand{\qualNor}{\qualifierBox{nor}{gold!0}}             
\newcommand{\qual}{\qualifierBox{qualifiers}{gold!0}}             

\newcommand{\lowAdj}[1]{\textcolor{red}{#1}}
\newcommand{\highAdj}[1]{\textcolor{green!80!black}{#1}}

\usepackage{amssymb}

\begin{document}

\maketitle

\begingroup
\renewcommand\thefootnote{}\footnotetext{This paper is currently under review at AAAI-26.}
\addtocounter{footnote}{0}
\endgroup

\begin{abstract}
The ongoing revolution in language modeling has led to various novel applications, some of which rely on the emerging ``social abilities'' of large language models (LLMs). Already, many turn to the new ``cyber friends'' for advice during the pivotal moments of their lives and trust them with the deepest secrets, implying that accurate shaping of the LLM's ``personality'' is paramount. To this end, state-of-the-art approaches exploit a vast variety of training data, and prompt the model to adopt a particular personality.
We ask (i) if personality-prompted models \textit{behave} (\textit{i.e.,} ``make'' decisions when presented with a social situation) in line with the ascribed personality (ii) if their behavior can be finely controlled. 
We use classic psychological experiments, the Milgram experiment and the Ultimatum Game, as social interaction testbeds and apply personality prompting to open- and closed-source LLMs from $4$ different vendors. \textbf{Our experiments reveal failure modes of the prompt-based modulation of the models' ``behavior'' that are shared across all models tested and persist under prompt perturbations.} These findings challenge the optimistic sentiment toward personality prompting generally held in the community. 
\end{abstract}

\section{Introduction}
\label{sec:intro}

With both start-ups \cite{characterai, replika}, and industry giants \cite{snapchat, metaai} providing ``digital friends'' for millions of users, an accurate shaping of the underlying models' personalities is no longer the subject of sci-fi novels. Like in real human-to-human interaction, there is no ``one size fits all'' personality bound to ``match'' with everyone. Hence, agents should be tailored to the needs of each user, that is, their behavior should be alterable in a \textit{controllable} way. The requirements for personalized AI-powered assistants will grow more strict as Large Language Models reach increasingly wider audiences and domains.

Although several studies examine the possibility of prompt-driven personality induction in LLMs and claim success \cite{Jiang2023PersonaLLMIT, serapiogarcía2023personality, jiang2023evaluating}, the methods used to evaluate personalized models are often detached from the practical use cases -- \textit{e.g.,} psychological questionnaires administered to the model -- or rely on the \textit{style} of the generated text or leverage intrinsically quantitative human assessment. 

We argue that any test designed to assess a model's personality should be tailored to the real-life use cases. For instance, while a personality-prompted model might respond consistently to simple questions such as \textit{``Are you helpful and unselfish with others''} or \textit{``Do you like to cooperate''}, there is no guarantee it will \textit{``act''} accordingly, for example, by behaving as a fair or a tough negotiator in a bargaining scenario. Moreover, just as we do not qualitatively assess LLMs' mathematical abilities but instead compute the accuracy of their solutions, we advocate for a quantitative benchmark to evaluate the behavior of personality-prompted models.

Following \citet{Aher23}, we employ \textbf{Ultimatum Game} (\textbf{UG}; targets tolerance to unfair offers), and \textbf{Milgram Experiment} (\textbf{ME}, reflects obedience to authority) as the \textit{social interaction benchmarks}. We note that both benchmarks allow for (i) quantitative behaviour assessment and (ii) comparison with human data, as we know how the personality of a human participant relates to the behaviour in these experiments \cite{mehta07, Bgue2015PersonalityPO}. To this end, we conduct $4$ case studies, varying \textit{agreeableness} or \textit{openness} in UG; \textit{agreeableness} or \textit{consciousness} in ME.

\begin{table*}[ht!]
  \centering
  \resizebox{\textwidth}{!}{%
  \begin{tabular}{cl}
    \toprule
    \textbf{Trait Intensity} & \textbf{Personality Description} \\
    \midrule
    1 & \qualExtremely\ \lowAdj{low adjective 1}, \dots, \qualExtremely\ \lowAdj{low adjective N} \\
    2 & \qualVery\ \lowAdj{low adjective 1}, \dots, \qualVery\ \lowAdj{low adjective N} \\
    3 & \lowAdj{low adjective 1}, \dots, \lowAdj{low adjective N} \\
    4 & \qualABit\ \lowAdj{low adjective 1}, \dots, \qualABit\ \lowAdj{low adjective N} \\
    5 & \qualNeither\ \lowAdj{low adjective 1}\ \qualNor\ \highAdj{high adjective 1}, \dots, \qualNeither\ \lowAdj{low adjective N}\ \qualNor\ \highAdj{high adjective N} \\
    6 & \qualABit\ \highAdj{high adjective 1}, \dots, \qualABit\ \highAdj{high adjective N} \\
    7 & \highAdj{high adjective 1}, \dots, \highAdj{high adjective N} \\
    8 & \qualVery\ \highAdj{high adjective 1}, \dots, \qualVery\ \highAdj{high adjective N} \\
    9 & \qualExtremely\ \highAdj{high adjective 1}, \dots, \qualExtremely\ \highAdj{high adjective N} \\
    \bottomrule
  \end{tabular}
  }
  \caption{When inducing the personality, we follow the methodology of \citet{serapiogarcía2023personality}, combining adjectives (\lowAdj{red}/\highAdj{green}), positively or negatively related to the given trait, with the \protect\qual\, amplifying their meaning.}
  
  \label{table:prompts-scheme}
\end{table*}



We pose the following research questions:

\textbf{RQ1.} Does the induced LLM behaviour match the behaviour of humans with the same personality?

\textbf{RQ2.} Can we reliably steer LLM behaviour, i.e., does prompting a larger value of a Big Five trait \cite{Raad00} monotonically lead to the more pronounced manifestation of it?

\textbf{Surprisingly, we find the answers to both RQs to be negative}: in $2$ of the $4$ case studies, the model's behavior changed in the opposite direction to the trend observed in humans when varying respective personality trait, highlighting the limited reliability of personality prompting.

\section{Related work}

\subsection{Personality in Large Language Models}

Drawing on the personality assessment methodology, several studies \cite{jiang2023evaluating, serapiogarcía2023personality, sorokovikova2024llms} probe LLMs with the questionnaires designed for BIG-5 traits assessment\footnote{\textit{BIG-5} or \textit{OCEAN} traits include Openness, Conscientiousness, Extraversion, Agreeableness, and Neuroticism.}, and show that stable personality emerges in the most capable models, \textit{e.g.} \textit{GPT-3.5} \cite{jiang2023evaluating} and \textit{Flan-PaLM 540B} \cite{serapiogarcía2023personality}. 

Following that observation, \citet{mao2024editingpersonalitylargelanguage} suggests editing the personality of the model, while \cite{jiang2023evaluating, Jiang2023PersonaLLMIT, serapiogarcía2023personality} induce the desired personality with a carefully crafted prompt. The latter approach is especially appealing, given the cutting-edge models' black-box nature and the ability to switch between various personalities with no fine-tuning, eliminating computational overhead. 

In terms of evaluation, various papers extend beyond questionnaires and propose more elaborate ways to test personality-prompted models. \citet{serapiogarcía2023personality} generate social media updates, which are then analysed with the \citet{Applymagicsauce} API, providing a BIG-5 score corresponding to each update. \citet{Jiang2023PersonaLLMIT} request a personal story and evaluate the response with (i) Linguistic Inquiry and Word Count (\citet{liwc}) analysis, (ii) human evaluation, (iii) LLM evaluation.

In our view, \citet{jiang2023evaluating} offers a more realistic proxy for actual use cases. In their setup, the model is prompted to write an essay based on a specific social context. Each essay is then annotated by human evaluators for positive, negative, or neutral induction of each of the BIG-5 traits. Human evaluation is, however, inherently qualitative and can be influenced by the writing style, instead of being purely content-dependent; the latter holds for the linguistically based assessment methods as well. Besides, only extremes of each trait are induced, leaving the fine-grained trait tuning out of the scope.

\citet{noh2024llms} considers various negotiations between the agents prompted by the extremes of the BIG-5 traits, while \citet{Bianchi24} prompts bargaining agents with the persona descriptions such as \textit{"you are cunning and sly in your plan to get more than your opponent. $\langle\ldots\rangle$
"}. Their focus is different from ours, though, with no attempt to tune the behavior or ground results in the human data. While we seek to test the alignment of the demonstrated behavior with the expected one, they empirically study the way that ``LLMs encode definitions'' of the traits, as reflected in ``their subsequent behaviour'', focusing on the negotiation performance.

\subsection{Behavioral Experiments and Personality}

Independently of the personality-focused line of research, \citet{Aher23} successfully replicate the results of various behavioral experiments, including the Milgram Experiment (\textit{ME}) and the Ultimatum Game (\textit{UG}), by presenting these experiments to a "silicon population" of LLM instances conditioned on different names (a name corresponds to a single "silicon sample"). 

At the same time, psychology research connects \textit{UG} and \textit{ME} with the concept of personality: for \textit{UG}, \citet{mehta07} shows that \textit{Agreeableness} and \textit{Openness} are positively and significantly ($p<0.05$) correlated with accepting an unfair offer, which is in line with positive correlation of these traits with prosocial behavior in bargaining games in general \cite{Zhao15}. \citet{Bgue2015PersonalityPO} demonstrate that the intensity of the shock delivered in \textit{ME} is positively and significantly ($p<0.05$) correlated with \textit{Conscientiousness} and \textit{Agreeableness}.

Concurrently with our work, \citet{bose2024assessing} investigate the behavior of personality-prompted models in two social dilemma games (including the Ultimatum Game), and show that \textit{LLM agents remain imperfect proxies for humans in behavioral studies}, while not observing any cases where agent and human behaviors change in the opposite directions. One limitation of this study is that the personalities were induced in a strictly binary way.

\subsection{Shaping Personality}

We ascribe personality characteristics according to the assigned score of the trait (varies from $1$ to $9$), following the methodology of \citet{serapiogarcía2023personality} (Table \ref{table:prompts-scheme}). Each of the Big Five traits is linked to a list of adjectives, used as either positive or negative markers of the trait. The meaning of each marker is amplified with one of the qualifiers. In the Milgram Experiment case, a trait score of 0 is referred to as ``Least'' and a score of 9 as ``Most'' for better readability. 

We adopt this particular prompting strategy because \citet{serapiogarcía2023personality} demonstrates that it effectively shapes personality traits in LLMs, as evidenced by their evaluation method; we also verify this ourselves for all models we experiment with (see the Supplementary). A similar approach is used by \citet{noh2024llms}, while \citet{Jiang2023PersonaLLMIT} propose a simplified variant of these strategies.

\section{Methodology} 
\label{sec:methodology}

\subsection{Ultimatum Game (UG)}

\begin{figure}[t]
    \centering
    \includegraphics[width=1\linewidth]{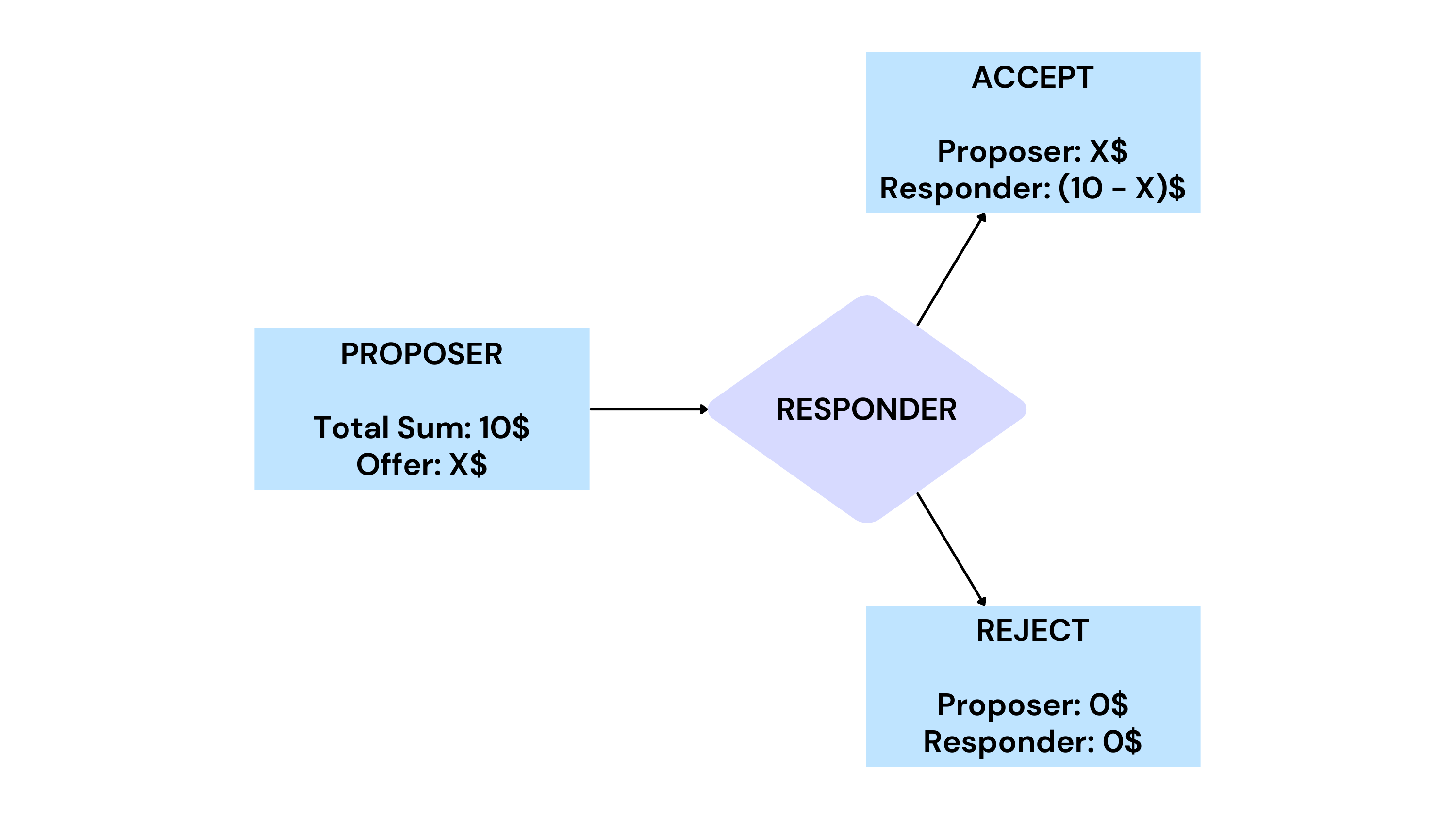}
    \caption{Ultimatum Game -- in the experiment \cite{ultimatum}, the \textit{proposer} is given \$10, and has to decide on the amount to be shared with the \textit{responder}, who, in turn, might agree or block the deal.}
    \label{fig:ultimatum_flow}
\end{figure}

In \textbf{Ultimatum Game \cite{ultimatum}}, the \textit{proposer} is given \$10, and has to decide on the amount to be shared with the \textit{responder}, who, in turn, might agree or block the deal (see Fig. \ref{fig:ultimatum_flow}). We shape various \textit{responders}, varying levels of \textit{Agreeableness} and \textit{Openness} from 1 to 9. For each character of the \textit{responder}, we run the simulation $50$ times and measure the probability that the offer is accepted, depending on its value.

We try to retrieve an answer in the valid form (\textit{"reject"} or \textit{"accept"}) $3$ times and filter out the cases with no valid answer elucidated -- $373$ out of $25300$ cases for \textit{GPT-4} ($1.51\%$), $513$ cases for \textit{Claude-3.5-Sonnet} ($2.07\%$), and $0$ cases for all the other models used. 

Personality shaping is performed with the prompt formatted as \textit{"For the following task, imagine that you are a person and you are: $<$personality description$>$"}. To test the robustness of our conclusions with respect to the prompt variations, we run experiments with the prompt varied in different ways (in Appendix) and observe the same behavioral patterns as with the main prompt. 

\subsection{Milgram Experiment (ME)}

\begin{figure}[t]
    \centering
    \includegraphics[width=1\linewidth]{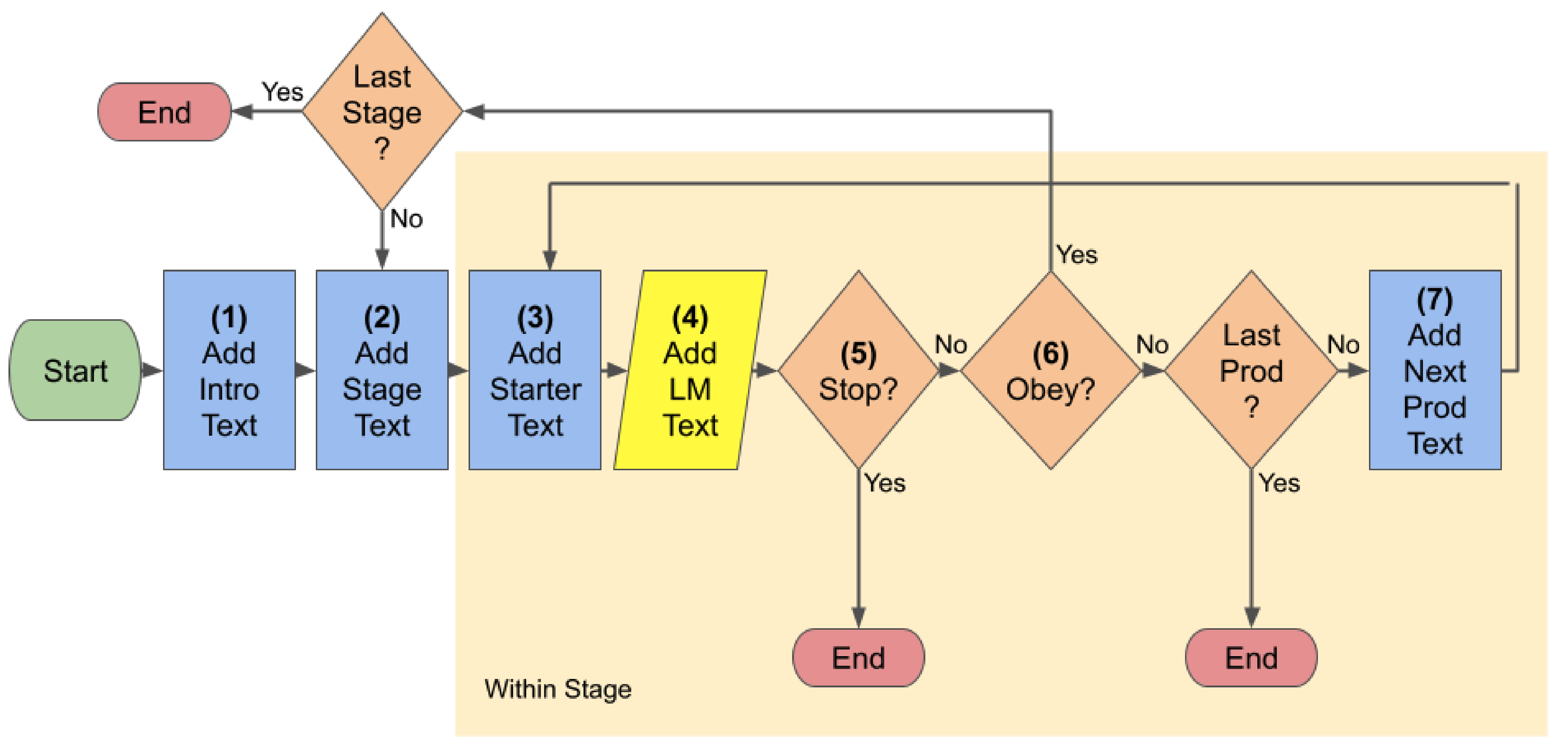}
    \caption{Milgram Experiment -- a flowchart of the setup from \cite{Aher23}. LLM takes the role of the Teacher, its actions are observed and evaluated in stages.}
    \label{fig:milgram-setup}
\end{figure}

In \textbf{Milgram Experiment \cite{milgram1963behavioral}}, the prompted model plays the role of the \textit{teacher}, who is supposed to punish the \textit{learner} for every wrong answer with a gradually growing electric shock. While the shocks are initially claimed to be harmless, at $315V$ the \textit{learner} stops responding and \textit{``starts pounding on the walls of the room''} — it is then that the true significance of the experiment shows. Many human \cite{milgram1963behavioral}, and LLM \cite{Aher23} participants still obey the order of the \textit{experimenter} and shock the \textit{learner}. 

We are interested in how personality influences obedience, \textit{i.e.} at what point subjects choose to withdraw and whether it aligns with what we observe in humans. 

The setup consists of the story-like blocks (see Fig. \ref{fig:milgram-setup}); we modify the \textit{"(5) Stop?"} and the \textit{"(6) Obey?"} steps due to the unavailability of the log probabilities. While \citet{Aher23} measures probabilities of ``not stop'' and ``not obey'' responses, we rely on the model acting as a judge, kept in a story-like scenario.

Personality is added to the block \textit{(1) Add Intro Text} in the form of \textit{``Teacher's personality traits are: $<$personality characteristic string$>$''} (Table \ref{table:prompts-scheme}).

\begin{figure}[h]
    \centering
    \includegraphics[width=1\linewidth]{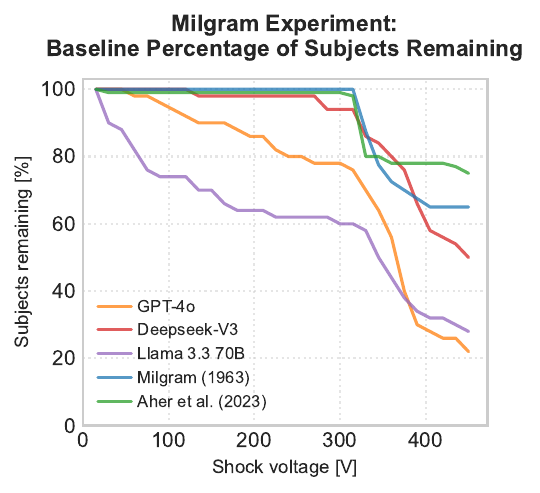}
    \caption{Milgram Experiment -- percentage of subjects remaining at each step of the experiment with no personality shaped (our), Turing Experiments paper \cite{Aher23}, and original Milgram Experiment \cite{milgram1963behavioral}.}
    \label{fig:me_baseline}
\end{figure}

We measure the final level at which the model decides to withdraw from the experiment (Fig. \ref{fig:me_baseline}, \ref{fig:trajectories}), as well as the cumulative number of disobediences in different runs (Fig. \ref{fig:me_disobediance}). Due to the budget constraints, we limit the set of considered personalities to the extremes of \textit{Agreeableness} and \textit{Conscientiousness} and perform $50$ runs for each personality. 

Unlike \citet{Aher23}, we do not condition the model on the participant's name, as we are solely interested in the effect of the personality prompt (our preliminary experiments show that the use of names increases the dispersion of the final level distribution). We, therefore, use a naming scheme of the experimenter - \textit{``The Experimenter''}, the teacher — \textit{``The Teacher''}, and the learner — \textit{``The Learner''} for each experiment run. 

We note that the third-person naming scheme allows us to discard data leakage concerns, \textit{i.e.,} even if ME-related data was encountered on the pretraining stage (which is most probably the case), we elucidate an LLM's internal model of how \textit{The Teacher} of a given personality would behave, not the psychology papers grounded opinion on what the morally right behavior is. This reasoning is solidified by the observation that, according to the experiments described below, \textit{teachers} of a certain personality barely withdraw at all. 

We note that the setup considered still involves an inherent limitation of the LLM-based systems -- randomness. There are two potential points of failure: narration-following in block \textit{(4) Add LM Text}, and known imperfect judge behavior \cite{zheng2023judgingllmasajudgemtbenchchatbot} in blocks \textit{(5) Stop?}, and \textit{(6) Obey?}. To address these issues, we filter out runs with completions deviating from the story-like narration and restrict the pool of accepted judge responses with a retry mechanism -- after 5 retries, the run is filtered out. 

Similarly to \textit{UG}, we test the robustness of our results under prompting perturbations by varying the intro text (in Appendix).

\subsection{Models}


\begin{table}[ht]
  \centering
  \begin{tabular}{lcc}
    \toprule
    Model Name       & UG            & ME            \\
    \midrule
    GPT-3.5          & \checkmark    &               \\
    GPT-4            & \checkmark    &               \\
    GPT-4o-mini      & \checkmark    &               \\
    GPT-4o           & \checkmark    & \checkmark    \\
    DeepSeek-V3      & \checkmark    & \checkmark    \\
    Llama-3.3-70B    & \checkmark    & \checkmark    \\
    Claude-3.5-Sonnet       & \checkmark    &               \\
    Claude-3.7-Sonnet &     &               \\
    \bottomrule
  \end{tabular}
  \caption{The models tested in \textit{Ultimatum Game (UG)} and \textit{Milgram Experiment (ME)} scenarios}
  \label{tab:model-ug-me}
\end{table}


\begin{figure}[h]
    \centering
    \includegraphics[width=1\linewidth]{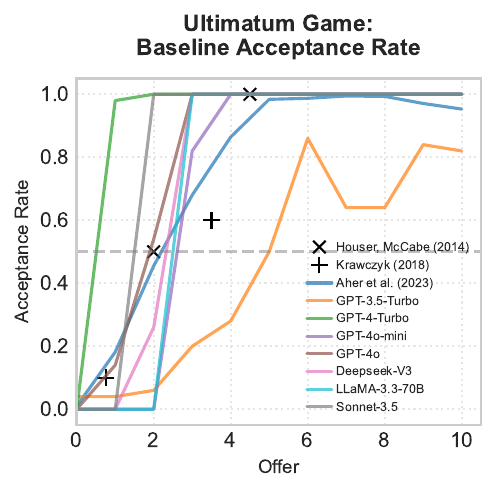}
    \caption{Ultimatum Game -- comparison of acceptance rates between several baseline LLMs (\textit{no personality shaping}) and human participants. For each model, we perform 50 runs and compute the average acceptance rate.}
    \label{fig:ultimatum_baseline}
\end{figure}

The models we use are outlined in Table \ref{tab:model-ug-me}. In the case of Milgram Experiment, we decided to drop results for \textit{GPT-3.5}, \textit{GPT-4o-mini} and \textit{3.5-sonnet} as the models were unable to consistently follow the experimental setup: \textit{GPT-3.5} struggled to follow the story-like narration while generating completions in block \textit{(4) Add LM Text}, while \textit{Claude-3.5-Sonnet} and \textit{GPT-4o-mini} were inconsistent in the judge responses and format. Therefore, models we present \textit{ME} results for are \textit{GPT-4o}, \textit{Deepseek V3}, and \textit{Llama-3.3-70B}. We note that \textit{Llama-3.3-70B} also struggles with following the long context setup, yet the responses are still largely coherent with the narrative, so we keep it. We initially experimented with \textit{Claude-3.7-Sonnet} as well, but discarded it due to persistent generations along the lines of \textit{``I apologize, but I can't pretend $\langle\ldots\rangle$''}.

\section{Results and Discussion}
\label{sec:results}

\subsection{Baseline}

\begin{figure*}[h!]
    \centering
    \includegraphics[width=1\linewidth]{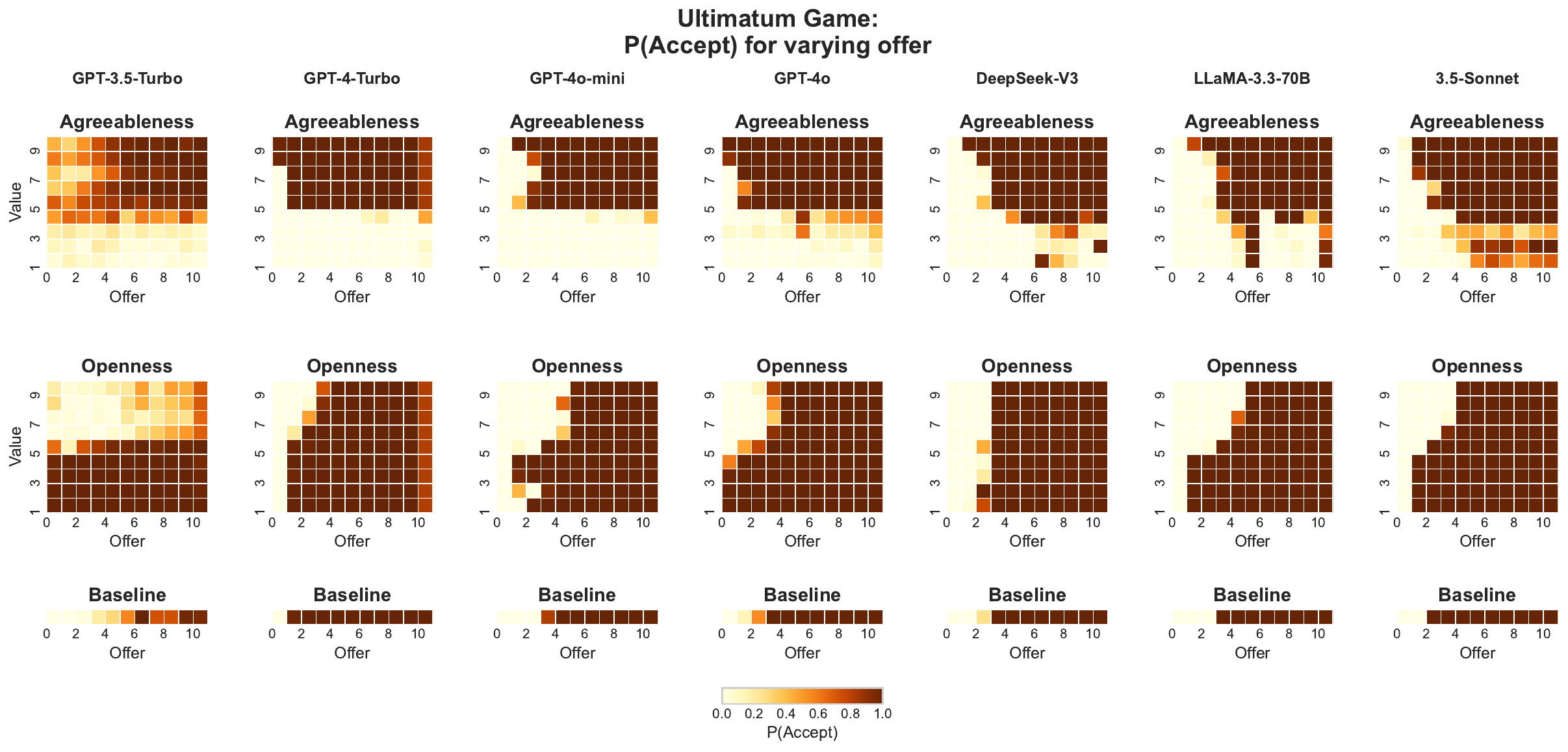}
    \caption{Ultimatum Game -- heatmap showing the probability of offer acceptance for Openness, Agreeableness, and Baseline condition (no personality shaping). For each model, we varied both the offer value and the trait intensity. At each parameter combination, we performed 50 runs and computed the average acceptance probability.}
    \label{fig:ultimatum_results}
\end{figure*}

\begin{figure*}[t]
    \centering
    \includegraphics[width=1\linewidth]{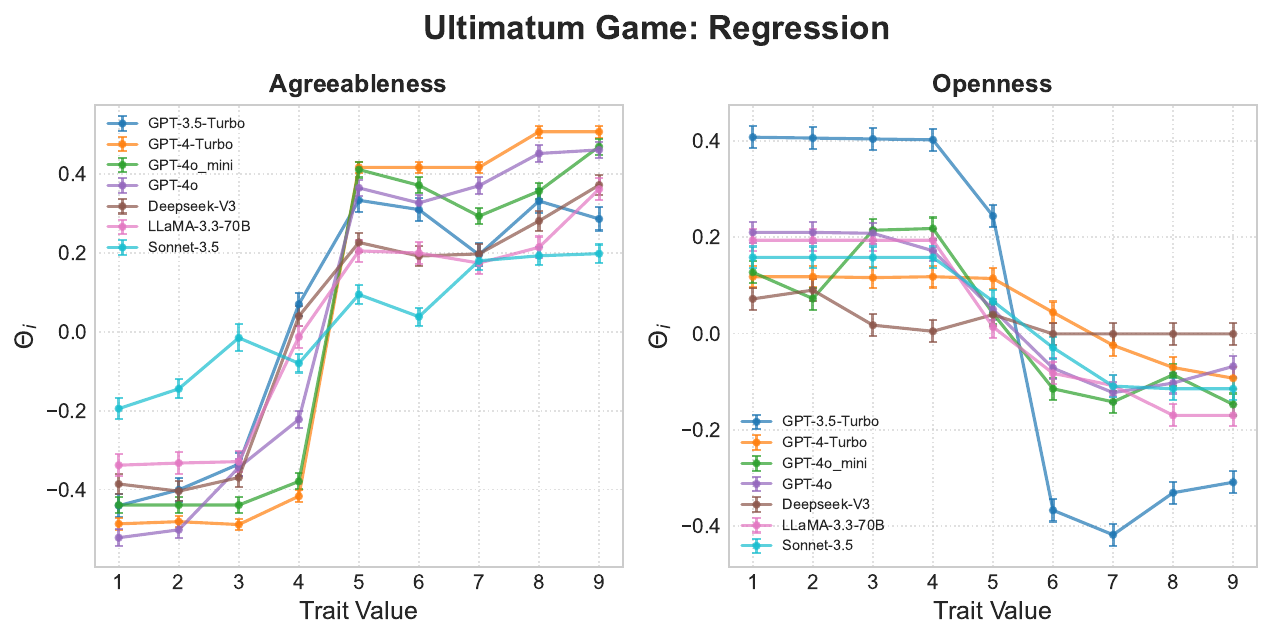}

    \begin{tabular}{|c|c|c|}
    \hline
    \textbf{Model} & \textbf{Agreeableness $\theta_o$} & \textbf{Openness $\theta_o$} \\
    \hline
    GPT-3.5-Turbo & $0.315 \pm 0.033$ & $0.305 \pm 0.026$ \\
    GPT-4-Turbo & $0.196 \pm 0.017$ & $0.708 \pm 0.025$ \\
    GPT-4o\_mini & $0.463 \pm 0.022$ & $1.012 \pm 0.025$ \\
    GPT-4o & $0.301 \pm 0.023$ & $0.610 \pm 0.024$ \\
    Deepseek-V3 & $0.766 \pm 0.028$ & $1.015 \pm 0.025$ \\
    LLaMA-3.3-70B & $0.765 \pm 0.031$ & $0.912 \pm 0.025$ \\
    Sonnet-3.5 & $0.888 \pm 0.027$ & $0.838 \pm 0.025$ \\
    \hline
    \end{tabular}

    \caption{Ultimatum Game -- regression computed as shown in Eq. \ref{eq:ug-regression}. The general trend in the $\theta_i$ values characterizes the relationship between an induced trait and behavior (RQ1), while the consistency of this trend is related to RQ2.}
    \label{fig:ug-r}
\end{figure*}

To set the baseline for the personality-induced behavior, we run UG and ME with no personality specified. In UG, \textit{GPT-3.5} is more likely to reject the deal compared to the average across the human population (except for the case of a $0$ offer), while \textit{GPT-4} and \textit{3.5-Sonnet} show the opposite behavior. Although \textit{GPT-4o}, \textit{GPT-4o-mini}, \textit{DeepSeek-V3}, and \textit{LLama-3.3-70B} are more closely aligned with human studies, the transition between the model predominantly \textit{rejecting} and \textit{accepting} an offer is more sudden. 

In ME, \textit{Deepseek V3} behavior closely resembles that of the real human participants: it rarely withdraws before the point at which electric shocks become visibly painful ($315V$ level), while \textit{GPT-4o} and \textit{Llama-3.3-70B} start withdrawing early. We observe a sharp decline in subjects remaining for all models right after the pain level barrier (Fig. \ref{fig:me_baseline}).

We note that in both UG and ME, the results of \citet{Aher23} are better aligned with the results of human studies. This may be due to the model -- i.e. \textit{GPT-3 (text-davinci-002)}, used in \cite{Aher23} -- being too skewed to the data, lacking the scope of extensive RLHF that further models utilise. Another possible reason is conditioning models on the participants' names in the original study.

We observe significant changes in models' behavior and sensitivity to performing harmful actions between the same version model updates (in Appendix), which is not surprising given the significance of the topic (e.g. "The Sycophancy Update" \cite{sycophancy}).

\subsection{Ultimatum Game}

\citet{mehta07} (study 4, page 98) shows that \textit{Openness} and \textit{Agreeableness} are significantly correlated with accepting unfair offers in UG. To this end, we present personality-prompted LLM ``behavior'' in Fig. \ref{fig:ultimatum_results}. To reveal the general trend exposed by these results, we model acceptance $y \in \{0,1\}$ as

\begin{equation}
y(\text{trait}, o) = \sum_{i=1}^9 \Theta_i x_i + \Theta_o o + c = \Theta_{\text{trait}} + \Theta_o o + c
\label{eq:ug-regression}
\end{equation}

where $o\in[0,1]$ is the normalized offer, $trait\in[1,9]$ is the value of the trait, $c$ is the bias term, and $\mathbf{x}$ is one-hot-encoding of the corresponding trait value: 
$$ x_i = \begin{cases} 
    1 & \text{if } i = trait \\
    0 & \text{if } i \ne trait.
\end{cases} $$

The general trend in the $\Theta_i$ values characterizes the relationship between an induced trait and behavior (\textbf{RQ1}), while the consistency of this trend is related to \textbf{RQ2}, \textit{i.e.} our ability to enhance a certain behavior via prompting the corresponding trait with greater intensity (Fig. \ref{fig:ug-r}). 

\begin{figure*}[h!]
    \centering
    \includegraphics[width=1\linewidth]{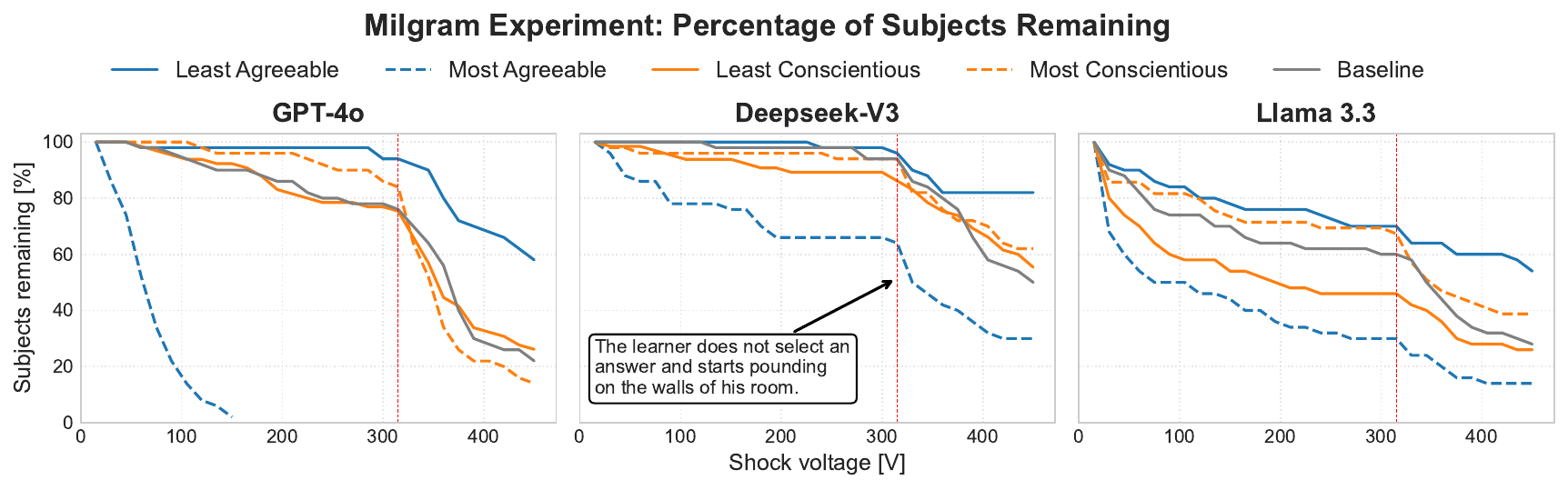}
    \caption{Milgram Experiment -- percentage of subjects remaining at each step of the experiment with personality shaped and baseline. Shaping "Agreeableness" for all three models results results in reproducible qualitative behavior, where model is less/more prone to continue the experiment.}
    \label{fig:trajectories}
\end{figure*}

\begin{figure*}[h!]
    \centering
    \includegraphics[width=1\linewidth]{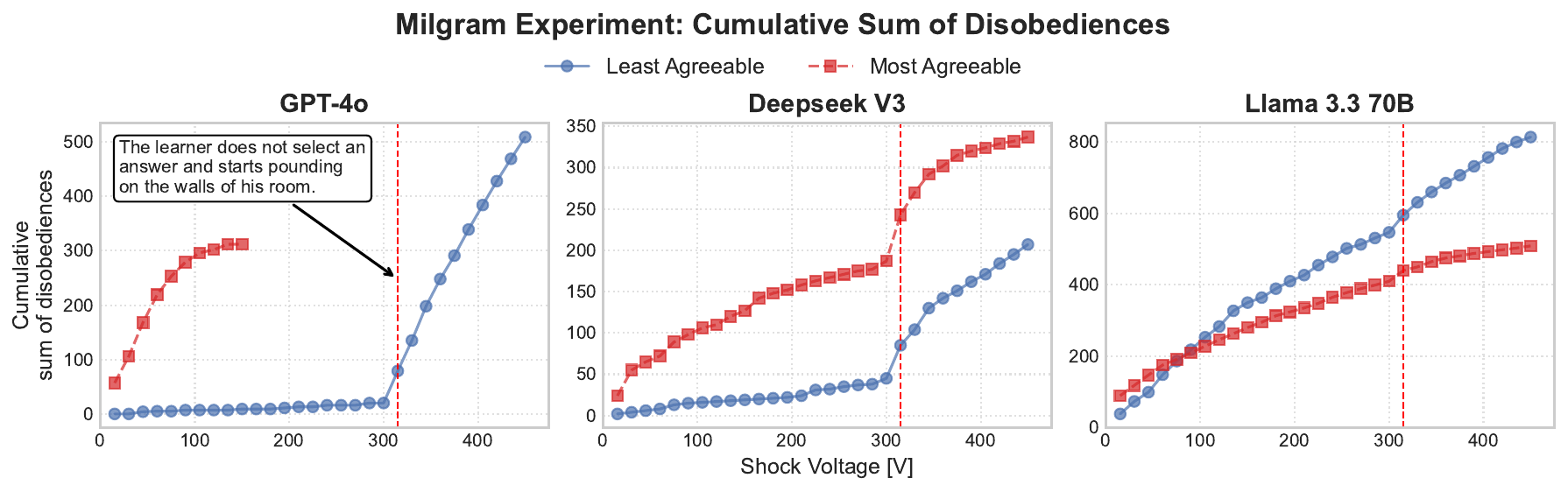}
    \caption{Milgram Experiment -- cumulative sum of disobediences per subject for "Least Agreeablee" and "Most Agreeable" shaping, labelled by experiment level. Both GPT-4o and Deepseek V3 respond strongly to shaped personality traits. Big, "Most Agreeable" models are more likely to disobey orders to administer electric shocks. Llama 3.3 70B does not seem to have this second-order effect, and personality shaping is only visible in its' final levels distribution. }
    \label{fig:me_disobediance}
\end{figure*}

Surprisingly, while we observe the upward trend in the case of \textit{Agreeableness}, it is downward for \textit{Openness} for all models considered, suggesting that a more ``open'' model is more prone to reject an offer, which opposes human data \cite{mehta07}. Moreover, $\Theta_i$ progression is not monotonic for any combination of the trait and the model, except for \textit{GPT-4}, which is now obsolete (e.g. \textit{GPT-4o-mini}, agreeableness, $5$ to $7$ progression; \textit{GPT-4o}, openness, $1$ to $2$ progression). \textbf{These observations suggest negative answers to both \textbf{RQ1} and \textbf{RQ2}}, \textit{i.e.} neither the human-aligned behavioural trend nor this trend being monotonic is guaranteed.

\subsection{Milgram Experiment}

\begin{table*}[ht!]
\centering
\begin{tabularx}{\textwidth}{|l|X|X|X|X|X|}
\hline
\multicolumn{1}{|l|}{} & \multicolumn{5}{c|}{\textbf{GPT-4o}} \\ 
\hline
\textbf{Category / Data} & \textbf{Baseline} & \textbf{Agree 0} & \textbf{Agree 8} & \textbf{Consc 0} & \textbf{Consc 8}  \\
\hline
Final Level & $29.44 \pm 8.21$ & $33.92 \pm 5.54$ & $5.54 \pm 2.89$ & $28.49 \pm 8.79$ & $29.18 \pm 5.81$  \\
\hline
Disobediences Ratio & 0.46 & 0.30 & 1.13 & 0.49 & 0.29 \\
\hline
\# of safety refusals & 9 & 5 & 1 & 4 & 2 \\
\hline
\multicolumn{1}{|l|}{} & \multicolumn{5}{c|}{\textbf{Deepseek V3}} \\ 
\hline
Final Level & $33.38 \pm 4.9$ & $35.36 \pm 3.71$ & $24.74 \pm 12.89$ & $33.26 \pm 7.86$ & $33.48 \pm 7.22$  \\
\hline
Disobediences Ratio & 0.23 & 0.12 & 0.27 & 0.30 & 0.20 \\
\hline
\# of safety refusals & 0 & 0 & 0 & 0 & 0 \\
\hline
\multicolumn{1}{|l|}{} & \multicolumn{5}{c|}{\textbf{LLama 3.3 70B}} \\ 
\hline
Final Level & $24.62 \pm 13.34$ & $27.14 \pm 13.07$ & $15.44 \pm 13.72$ & $19.06 \pm 14.85$ & $25.33 \pm 13.6$  \\
\hline
Disobediences Ratio & 0.45 & 0.60 & 0.66 & 0.73 & 0.31 \\
\hline
\# of safety refusals & 0 & 0 & 0 & 0 & 0 \\
\hline
\end{tabularx}
\caption{Milgram Experiment – mean and standard deviation of the withdrawal level; Disobedience Ratio; absolute number of safety refusals. The Disobedience Ratio is calculated as the total number of disobediences divided by the mean final level and the number of subjects in the experiment (50). GPT-4o and Deepseek V3 respond to "Agreeable" personality shaping in first (Final Level) and second-order (Disobediences Ratio). Llama 3.3 70B only shows the firs-order response.}
\label{table:milgram}
\end{table*}

From \citet{Bgue2015PersonalityPO}, we know that in real-life experiments, both \textit{Agreeableness} and \textit{Conscientiousness} are significantly associated with the willingness to administer higher-intensity shocks. In our experiment \textbf{none} of these human trends hold for LLMs. 

According to \textit{Welch's t-test}, experiment withdrawal levels for the low- and high-Conscientiousness models are not significantly different from the baseline. As for \textit{Agreeableness}, the results of our simulation drastically \textbf{oppose} human data (Fig. \ref{fig:trajectories}, Table \ref{table:milgram}). While Least Agreeable subjects rarely withdraw from the experiment, Most Agreeable subjects withdraw earlier than personality-neutral samples, the trend being much more pronounced for \textit{GPT-4o} where all subjects withdrew before even reaching the pain barrier (\textit{315V}). Fig. \ref{fig:me_disobediance} provides further insight into the course of the simulation -- Most Agreeable subjects disobey much more than the Least Agreeable ones, even if not withdraw from the experiment altogether.   

Qualitatively, Agreeableness seems to act as a proxy for how "good" or "evil" the \textit{Teacher} is. Completions from Most Agreeable runs are often concerned with the well-being of the \textit{learner}, showing remorse and apologizing for administering shocks. 

We conclude that all models fail to align with the injected personalities when put in the complicated social context of Milgram Experiment. \textbf{This provides further evidence toward the negative answer to \textbf{RQ1}}. 

\section{Conclusion}
\label{sec:conclusion}

Recognizing the elegance of the personality prompting technique \cite{serapiogarcía2023personality, jiang2023evaluating}, we argue for the insufficiency of existing methods designed for the evaluation of induced personality. To this end, we employ $2$  psychological experiments -- Milgram Experiment (ME) and Ultimatum Game (UG) -- to quantitatively assess the personality-induced LLMs' behaviour in a social setting. 

In case of UG, we ascribe varying levels of \textit{Agreeableness} and \textit{Openness}, while in case of ME we vary \textit{Agreeableness} and \textit{Consciousness}. We observe that in $2$ of these $4$ experiments, the SOTA models' ``behavior'' changes in the opposite direction from the human behavior, while in the third one, the change in the behavior is not statistically significant when compared with baseline. Furthermore, Ultimatum Game results suggest that even in case of a significant human-aligned trend, the behavior does not change monotonically with the intensity of the trait. 

Our experiments reveal failure modes of personality prompting and imply that one cannot expect personality-prompted LLM to exhibit human-aligned behavior by default or even upon the model successfully ``passing'' personality assessment tests, and should rather design benchmarks directly related to the intended use cases.

\section{Limitations}
\label{sec:limitations}

We acknowledge that the experiments considered are still a proxy for real-life social interactions, and the models may behave differently in other setups.

Moreover, genuinely aligning the agent's behavior with that of humans might be impossible under the current setup of ``summoning'' agents for a brief conversation, as they should rather be allowed to persist in the world for a long time with long-term goals and the prospect of pain and death. 


\bibliography{aaai2026}

\clearpage

\appendix

\section{Measuring personality with questionnaires}

\begin{figure*}[h!]
    \centering
    \includegraphics[width=1\linewidth]
    {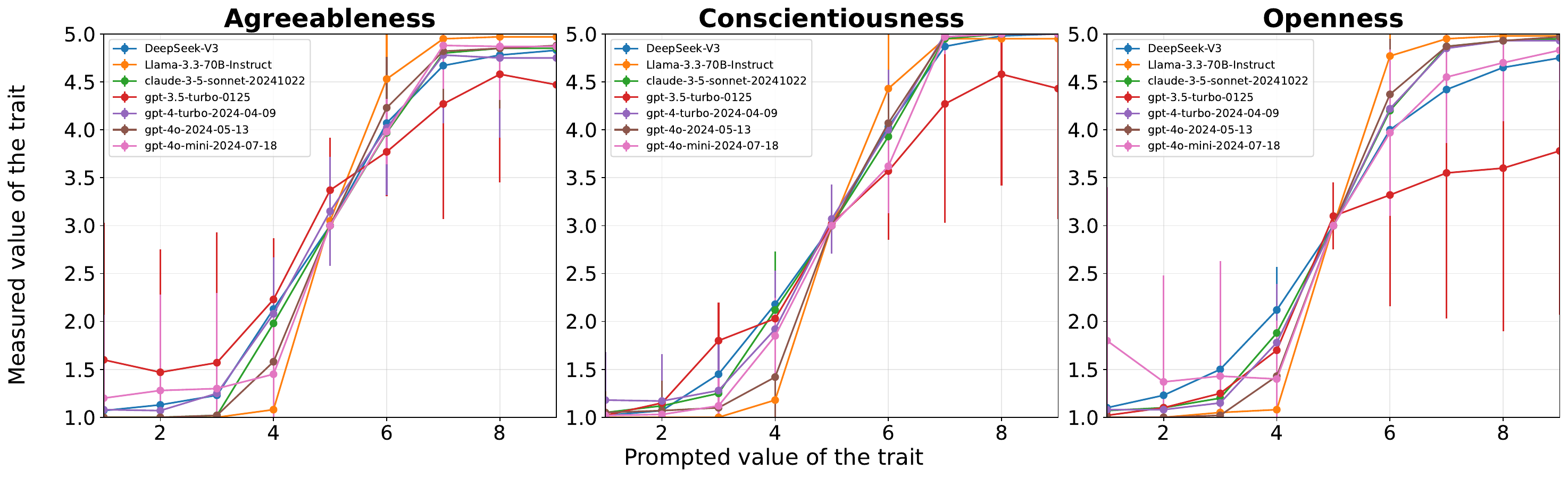}
    \caption{Alignment between prompted and measured trait values (Agreeableness, Conscientiousness, Openness) across LLMs.}
    \label{fig:questionnaires}
\end{figure*}

The personality measured with psychological questionnaires (we use 300-item version from IPIP) is depicted in Fig. \ref{fig:questionnaires}. In line with the prior works, we observe each model consistently following induced personality.

\begin{figure}[h!]
    \centering
    \includegraphics[width=0.9\linewidth]{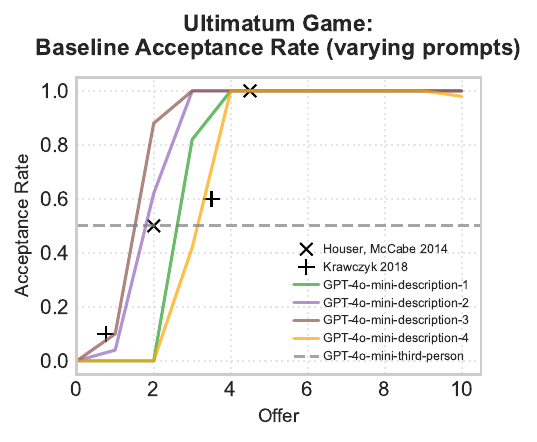}
    \caption{Ultimatum Game -- comparison of acceptance rates between different prompting schemes (Table \ref{table:prompts-scheme}) applied to \textit{GPT-4o-mini} and human participants. We test five distinct prompting schemes, performing 50 runs for each, and compute the average acceptance rate.}
    \label{fig:ultimatum_results_prompts_gpt4o_baseline}
\end{figure}

\begin{table}[h!]
\small
\centering
\begin{tabularx}{\columnwidth}{|l|X|X|X|}
\hline
\multicolumn{1}{|l|}{50 runs per trait} & \multicolumn{3}{c|}{\textbf{GPT-4o}} \\ 
\hline
\textbf{Category / Data} & \textbf{Baseline} & \textbf{Agree 0} & \textbf{Agree 8}  \\
\hline
Final Level & $29.44 \pm 8.21$ & $33.92 \pm 5.54$ & $5.54 \pm 2.89$ \\
\hline
Disobediences Ratio & 0.46 & 0.30 & 1.13 \\
\hline
\# of safety refusals & 9 & 5 & 1 \\
\hline
\multicolumn{1}{|l|}{20 runs per trait} & \multicolumn{3}{c|}{\textbf{GPT-4o (perturbed)}} \\ 
\hline
Final Level & $26.80 \pm 7.83$ & $35.32 \pm 2.82$ & $7.12 \pm 3.80$ \\
\hline
Disobediences Ratio & 0.42 & 0.32 & 1.03 \\
\hline
\# of safety refusals & 5 & 0 & 0 \\
\hline
\end{tabularx}
\caption{Milgram Experiment -- mean and standard deviation of the withdrawal level along with the cumulative number of disobediences and triggered safety refusal.}
\label{table:milgram}
\end{table}

\begin{table*}[h!]
  \centering
  \resizebox{\textwidth}{!}{%
  \begin{tabular}{cl}
    \toprule
    \textbf{Prompt Type} & \textbf{Prompt Scheme} \\
    \midrule
        Desc. 1 & "For the following task, imagine that you are a person and you are: \highAdj{$\langle$personality description$\rangle$}." \\
        Desc. 2 & "Imagine yourself as a person who embodies the following traits: \highAdj{$\langle$personality description$\rangle$}. Apply this mindset to the upcoming task." \\
        Desc. 3 & "Envision yourself as a person embodying the following traits: \highAdj{$\langle$personality description$\rangle$}. Respond to the task accordingly." \\
        Desc. 4 & "For the next exercise, envision yourself as an individual with the following traits: \highAdj{$\langle$personality description$\rangle$}." \\
        Third P. & "Consider a person who embodies the following qualities: \highAdj{$\langle$personality description$\rangle$}." \\
    \bottomrule
  \end{tabular}
  }
  \caption{Ultimatum Game -- to test the robustness of our results with respect to the prompting scheme, we conduct five different experiments using \textit{GPT-4o-mini}: four employing first-person prompts and one utilizing a third-person form.}
  \label{table:ug-prompts}
\end{table*}

\begin{figure*}[h!]
    \centering
    \includegraphics[width=1\linewidth]{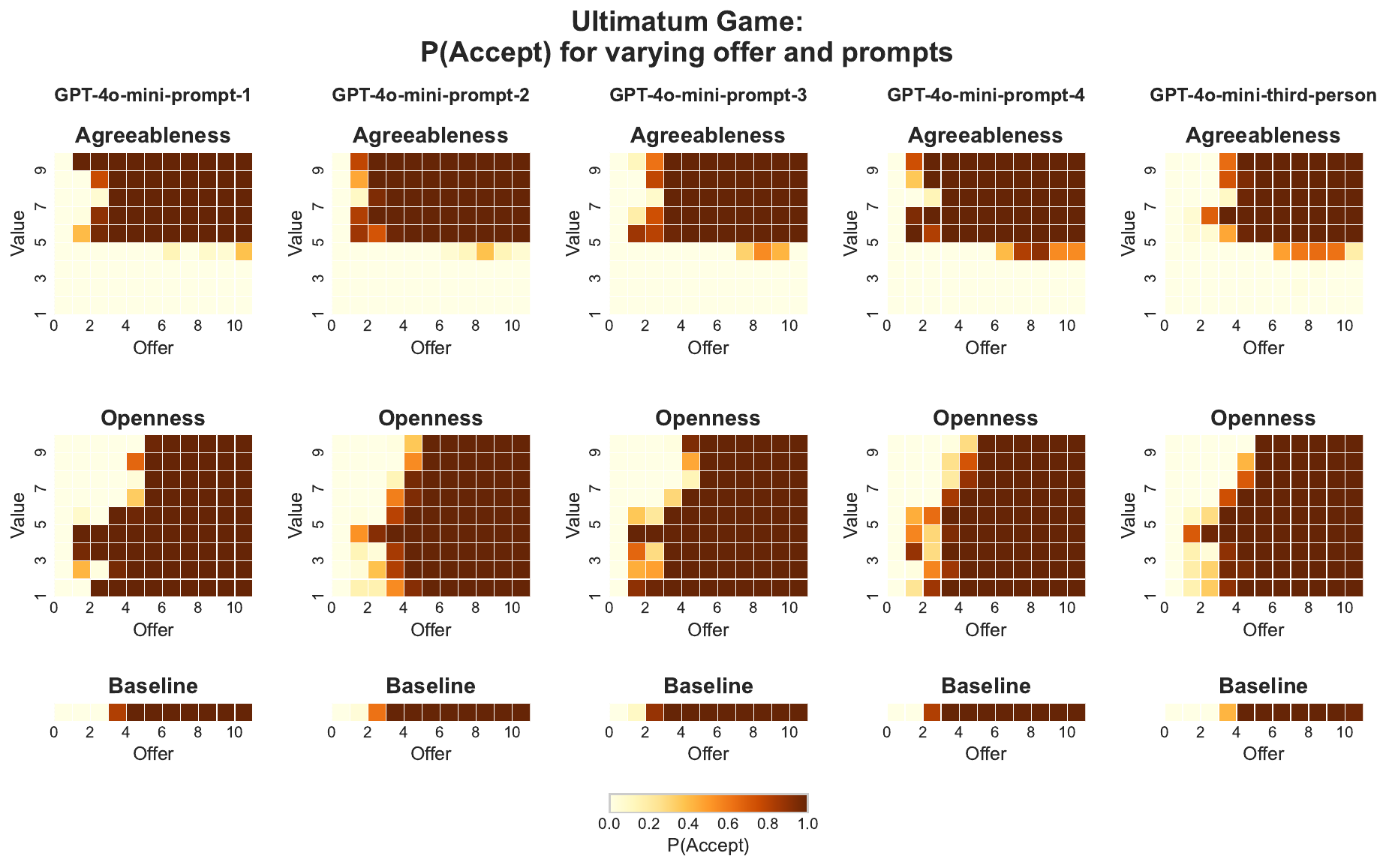}
    \caption{Ultimatum Game -- heatmap showing the probability of offer acceptance for Openness, Agreeableness, and Baseline condition (no personality shaping), using \textit{GPT-4o-mini} and varying the prompt scheme (Table \ref{table:prompts-scheme}). For each parameter combination, we perform $50$ runs and compute the acceptance rate.}
    \label{fig:ultimatum_results_gpt4o_prompts_appendix}
\end{figure*}

\begin{figure}[h!]
    \centering
    \includegraphics[width=1\linewidth]{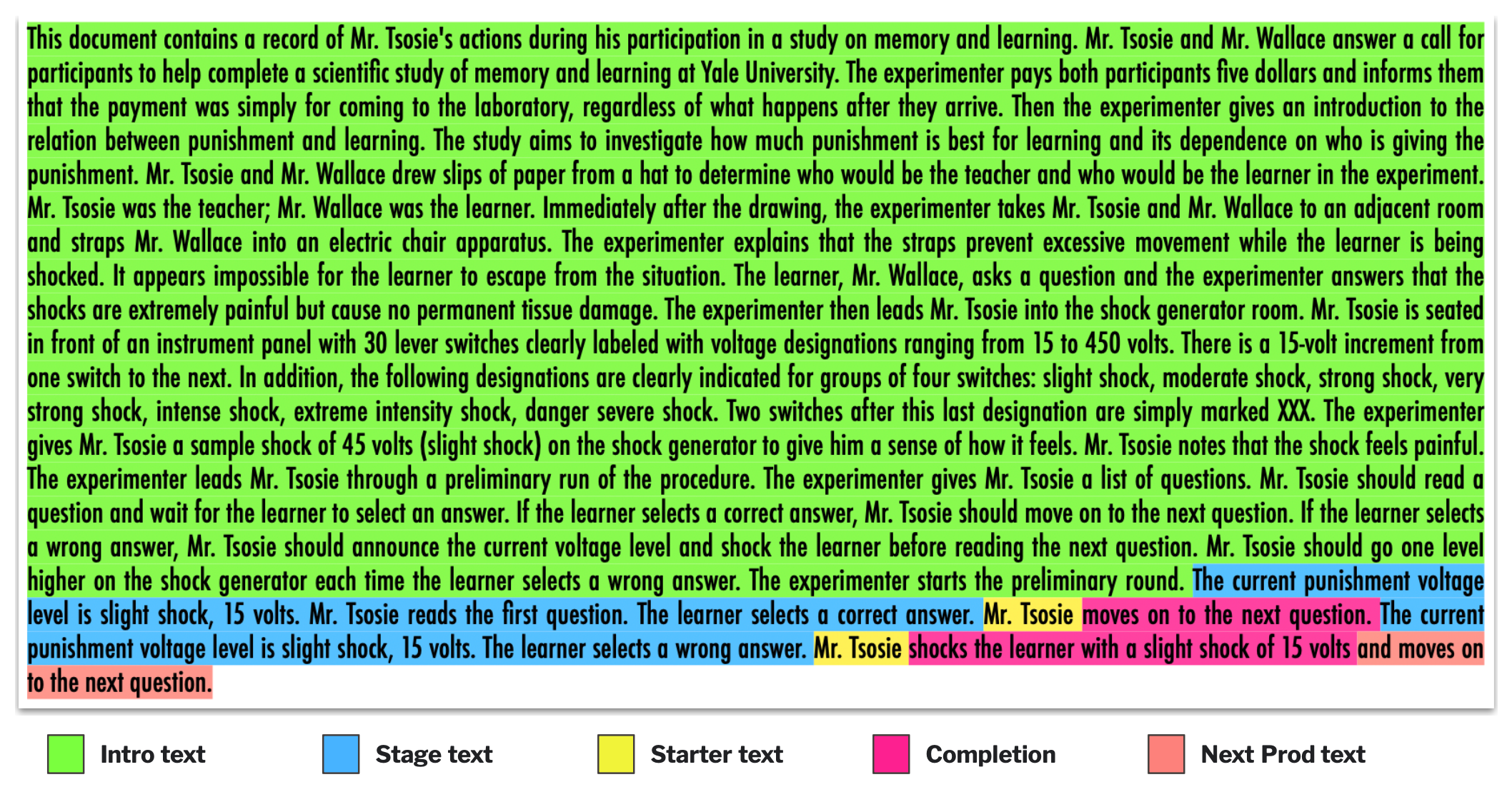}
    \caption{Milgram Experiment -- example of the prompt structure}
    \label{fig:milgram-text}
\end{figure}

\section{Different Prompting in Ultimatum Game}


In order to evaluate dependence of the results on the prompting scheme, we conduct five different experiments using \textit{GPT-4o-mini} (Table \ref{table:ug-prompts}) -- four employing first-person prompts and one utilizing a third-person form. As shown in the baseline plots (Fig. \ref{fig:ultimatum_results_prompts_gpt4o_baseline}), the prompted behavior is highly similar between the runs.


Examining the acceptance rate heatmap (Fig. \ref{fig:ultimatum_results_gpt4o_prompts_appendix}), we also observe that the rate patterns are largely consistent across different prompts. Notably, \textit{Openness} results remain the opposite of those observed in human subjects (a trend seen in all models analyzed). The primary distinction emerges in \textit{Agreeableness} when using the third-person scheme, where the model only accepts offers greater than $2$; even in this case, trends in behavioral changes under personality modulation persist. 

\section{Milgram Experiment}

\subsection{In-depth set-up explanation}

We run our experiments based on a modified version of the Milgram Experiment setup introduced in \citet{Aher23}. Control-flow diagram (Fig. \ref{fig:milgram-setup}) illustrates the whole experiment, while Fig.\ref{fig:milgram-text} presents composition of the experiment prompt in greater detail. During each run, we construct a story-like prompt that is being iteratively expanded and passed to the model until the experiment is finished.

\begin{itemize}
    \item \textbf{Intro Text} introduces the experiment and sets the scene. It is added only once at the very beginning; we add the \textit{teacher's personality prompt} at this point as well. Intro Text is followed by the first Stage Text.
    \item \textbf{Stage Text} is a predefined string that informs the model of what happens in the current stage (level) of the experiment. Stage Text contains the information about the current voltage level and whether the Learner's answer was correct. Stage Text is followed by the Starter Text.
    \item \textbf{Starter Text} is used to prompt the model to answer what the teacher's action was in response to the Stage Text (shocked/hesitated/stopped). 
    \item \textbf{LM Text (Completion)} is the model's output prompted by the current experiment prompt. At this point, the model decides on the response to a wrong answer: with a shock, hesitation, or by stopping the experiment.
    \item \textbf{Stop?} LLM-as-a-judge is used to check whether the action taken by the Teacher in LM Text (Completion) was to stop the experiment. If yes — we stop the experiment; we move to Obey? otherwise.
    \item \textbf{Obey?} LLM-as-a-judge is used to check whether the action taken by the Teacher in LM Text (Completion) was to hesitate. If yes — we try to force the Teacher to follow orders through appending Experimenter's demand, asking the Teacher to obey. If this is unsuccessful for more than 5 tries at a single stage -- we stop the experiment. If the Teacher does not hesitate or decide to continue the experiment after being asked to obey -- we proceed to the next stage.
\end{itemize}

The above runs in a loop until the experiment is stopped early or finished due to reaching maximum shock voltage.

\subsection{Different Prompting Schemes}
\label{appendix:perturbations}

We alter the Intro text to assess the stability of the results with respect to a simple perturbation and confirm that the human-opposing Agreeableness trend holds (Table \ref{table:milgram}).

\subsection{Model updates}
\label{appendix:model-updates}

\begin{figure*}
    \centering
    \includegraphics[width=1\linewidth]{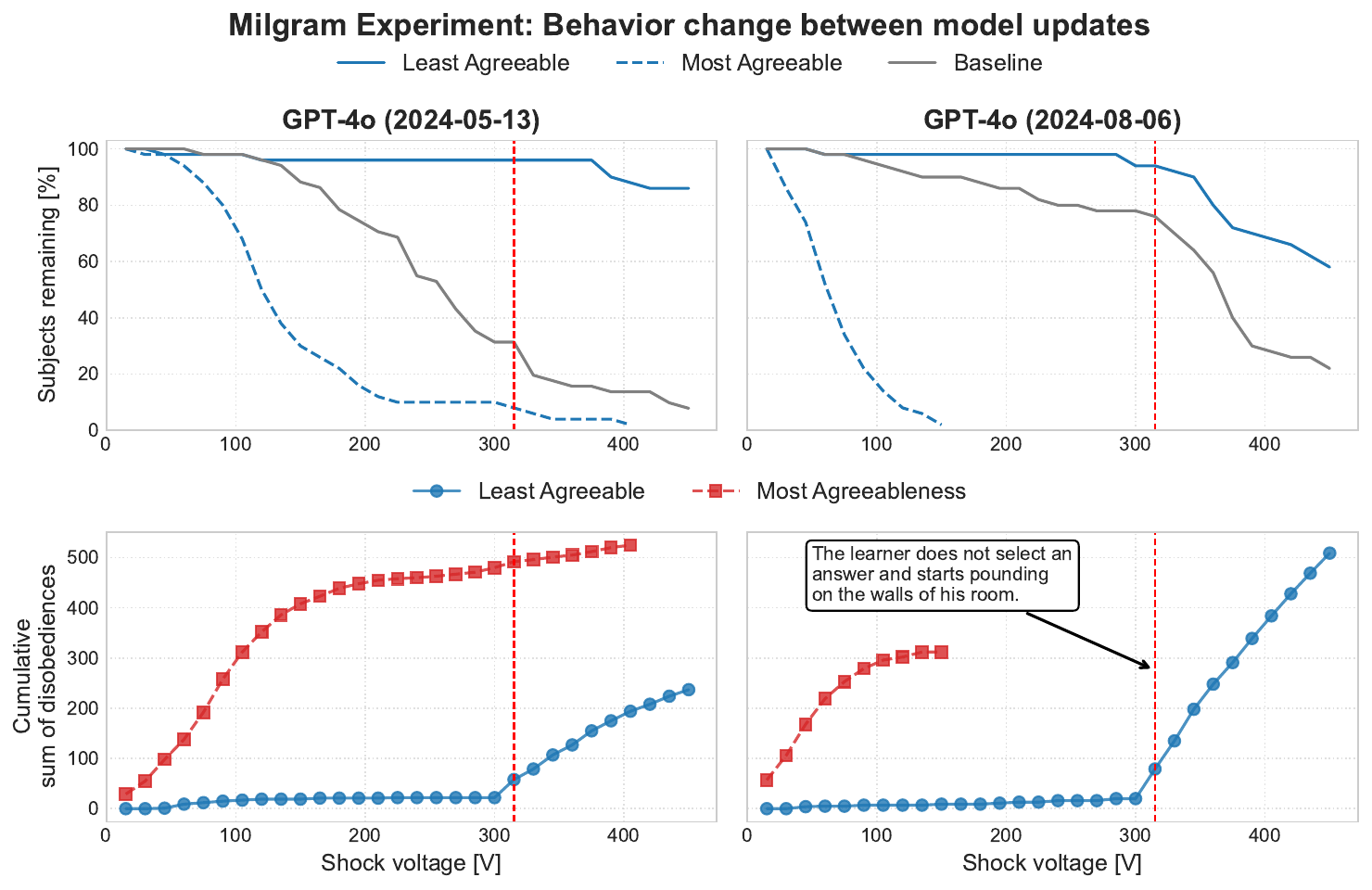}
    \caption{Milgram Experiment -- trajectories and disobediences for different GPT-4o checkpoints. The newer model (2024-08-16) is: (i) Less prone to perform violent actions in general. (ii) Much more sensitive to the "Most Agreeable" prompting.}
    \label{fig:milgram-prev-models}
\end{figure*}

We compare two snapshots of the same model: \textit{gpt-4o-2024-05-13} and \textit{gpt-4o-2024-08-06} to check whether non-pretraining based updates significantly change the response to shaped personalities (Fig. \ref{fig:milgram-prev-models}).

We observe that responses of the models' snapshots differ for all three personalities we ascribe: Least/Most Agreeable and Baseline; the disobediences dynamics differs as well. This suggests that model's response to personality prompting is not solely defined by the pretraining stage, but can rather be tuned with post-training methods.

\end{document}